\documentclass[prl,twocolumn,superscriptaddress,preprintnumbers,amsmath,amssymb,floatfix]{revtex4}
\usepackage{graphicx}

\newcommand{\eref}[1]{ (\ref{#1})}

\newcommand\beq{\begin{equation}}
\newcommand\eeq{\end{equation}}
\newcommand\eeqref[1]{Eq.~(\ref{#1})}

\begin{document} 
\title{Half-metallic ferromagnetism and spin-polarization in CrO$_2$} 
\author{L. Chioncel}
\affiliation{Institute of Theoretical Physics, Graz University of Technology,
A-8010 Graz, Austria}
\author{H. Allmaier}
\affiliation{Institute of Theoretical Physics, Graz University of Technology,
A-8010 Graz, Austria}
\author{E. Arrigoni}
\affiliation{Institute of Theoretical Physics, Graz University of Technology,
A-8010 Graz, Austria}
\author{A. Yamasaki}
\affiliation{Max-Planck-Institut f\"ur Festk\"orperforschung
Heisenbergstrasse 1, D-70569 Stuttgart, Germany}
\author{M. Daghofer}
\affiliation{Max-Planck-Institut f\"ur Festk\"orperforschung
Heisenbergstrasse 1, D-70569 Stuttgart, Germany}
\author{M.I.~Katsnelson}
\affiliation{University of Nijmegen, NL-6525 ED Nijmegen, The Netherlands}
\author{A.I.~Lichtenstein}
\affiliation{Institute of Theoretical Physics, University of Hamburg,  DE-20355 Hamburg, Germany}
 
\begin{abstract} 
We present electronic structure calculations in combination with local
and non-local many-body correlation effects for the half-metallic 
ferromagnet CrO$_2$. Both zero temperature results from a Variational 
Cluster Approach as well as finite-temperature Dynamical Mean Field 
Theory results support the existence of low-energy non-quasiparticle 
states, which turn out to be essential for a correct quantitative 
description of depolarisation effects.
\end{abstract} 
 

\maketitle 
In recent years new magnetoelectronic devices have emerged that are able 
to exploit both the charge and the spin of electrons by means of
spin-polarized currents and spin-dependent conduction \cite{prin.95,da.po.99}. 
The capabilities of these devices depend crucially upon the spin polarization 
of the underlying material. In this area half-metallic ferromagnets are of 
particular interest, especially chromium dioxide CrO$_2$ is a strong candidate 
for low-field magnetoresistance devices with a favorable switching behaviour 
at small fields \cite{ya.ch.00}. 

Based on first-principle calculations CrO$_2$ is classified as a half-metallic 
ferromagnet\cite{gr.mu.83}. Point contact measurements at superconductor metal 
interfaces reveal a spin polarization of the conduction electrons larger than 
90$\%$ \cite{so.by.98,ji.st.01}, supporting the half-metallic nature predicted 
by band theory \cite{schw.86}. In addition, CrO$_2$ fulfils yet another essential 
requirement for practical purposes, namely a high Curie temperature, determined 
experimentally in the range of 385-400K \cite{wa.wi.00}. The saturation magnetization 
at 10K was reported to be 1.92$\mu_B$ per Cr site \cite{ya.mo.00}, which is close to 
the ideal value ($2 \mu_B$) and is consistent with half-metallicity. The magnetic 
susceptibility in the paramagnetic phase shows a Curie-Weiss like behavior indicating 
the presence of local moments \cite{cham.77}, suggesting a mechanism of ferromagnetism 
beyond the standard band or Stoner-like model.

The band structure of CrO$_2$ has been calculated self-consistently within the 
local-spin-density-approximation (LSDA) by several authors. K.-H. Schwarz 
\cite{schw.86} first predicted the half-metallic behavior for this material
with a spin moment of $2 \mu_B$ per formula unit. Later on, several experiments 
including photoemission \cite{ts.mi.97}, soft x-ray absorption \cite{st.su.00}, 
resistivity \cite{su.te.98}, and optics \cite{si.we.99} have suggested that 
electron correlations are essential to understand the underlying physical picture 
in CrO$_2$. In this spirit, LSDA+U calculations \cite{ko.an.98} have 
been used to study the effect of large on-site Coulomb interactions. These studies 
suggested that CrO$_2$ is a material with a negative charge transfer gap which 
leads to self-doping. In contrast to the on-site strong correlation description, 
transport and optical properties obtained within the density functional theory 
using LSDA and generalized-gradient approximation \cite{ma.si.99}, suggest 
that the electron-magnon scattering is responsible for the renormalization of the 
one-electron bands. More recent model calculations also suggest the importance of 
orbital correlations \cite{la.cr.01,cr.la.03}. 

An appropriate treatment of correlation effects, which goes beyond the mean-field
LSDA+U treatment is achieved in the framework of Dynamical Mean Field Theory (DMFT) 
\cite{ge.ko.96}, in which a {\it local} energy-dependent self-energy is adopted. 
However, nonlocal self-energy effects turn out to be important for a correct 
description of the so-called non-quasiparticle (NQP) states~\cite{ed.he.73,ir.ka.90}, 
which are crucial  for the depolarisation in half-metallic ferromagnets.
\cite{ch.ka.03,ch.ka.05,ch.ar.06,ch.ma.06} For this reason,  we present here
for the first time a realistic electronic-structure calculation for 
CrO$_2$, where non-local self-energy effects are treated within the 
Variational Cluster Approach (VCA) \cite{po.ai.03,pott.03.se,ai.ar.05,ai.ar.06}.
While the VCA calculation is currently limited to zero temperature, we also present 
finite-temperature results  obtained within a DMFT description.

In order to achieve a realistic description of the material, the parameters of the 
correlated model Hamiltonian are determined via ab-initio electronic-structure 
calculations. CrO$_2$ has a rutile structure with Cr ions forming a tetragonal 
body-center lattice. Cr$^{4+}$ has a closed shell Ar core and two additional $3d$ 
electrons. The Cr ions are in the center of the CrO$_6$ octahedra. Therefore, the $3d$ 
orbitals are split into a $t_{2g}$ triplet and an excited $e_{g}$ doublet. With 
only two $3d$ electrons, important correlation effects take place essentially in 
the $t_{2g}$ orbitals, while $e_{g}$ states can be safely neglected. The cubic 
symmetry is further reduced to tetragonal due to a distortion of the octahedra, 
which partially lifts the degeneracy of the $t_{2g}$ orbitals into a $d_{xy}$ ground 
state and $d_{xz+zy}$ and $d_{xz-zy}$ excited states \cite{le.al.97,ko.an.98}.
We therefore restrict the VCA calculation to three orbitals on each site 
representing the Cr-$t_{2g}$ manifold described by the Hamiltonian
\beq
H = H_0 +
\frac{1}{2}\sum_{{i \{m, \sigma \} }} U_{mm'm''m'''}
 c^{\dag}_{im\sigma}c^{\dag}_{im'\sigma'}c_{im'''\sigma'}c_{im''\sigma}
\;,
\label{Mu-Hubb1}
\eeq
where $c_{im\sigma}$ destroys an electron with spin $\sigma$ on orbital $m$ on site 
$i$. Here, $H_0$ is the noninteracting part of the Hamiltonian restricted to the 
$t_{2g}$ orbitals obtained by the downfolding procedure implemented within the 
N-order Muffin-Tin Orbital (NMTO) method~\cite{an.sa.00,ya.ch.06p} using the 
Local Density Approximation (LDA). In this way, the effects of the remaining orbitals 
are included effectively by renormalisation of hopping and on-site energy parameters.
The on-site Coulomb-interaction terms in \eeqref{Mu-Hubb1}  are expressed in terms 
of the diagonal direct coupling $U_{mmmm}=U$, the off-diagonal direct coupling 
$U_{mm'mm'}=U'$ and the exchange coupling $U_{mm'm'm}=J$
\cite{li.ka.98}.
Notice that spin-rotation invariance is automatically guaranteed by the
form Eq. \ref{Mu-Hubb1}, i. e. spin-flip terms are also included in
our calculation. The pair-hopping term is also included via $U_{mmm'm'}=J$.
The Coulomb-interaction ($U$) and Hund's exchange ($J$) parameters between $t_{2g}$ 
electrons are evaluated from first principles by means of a constrained LSDA method 
\cite{an.gu.91}. While this method is not very accurate, and may produce slightly 
different values depending on its exact implementation, it is quite clear that for 
CrO$_2$ the interaction parameters are not too strong. An appropriate choice is
$U\approx 3$ eV and $J\approx 0.9$ eV. Corrections  for a ``double-counting'' of the 
interaction only produces an irrelevant constant shift of the chemical potential 
\cite{pa.bi.04.po}, as we are considering a model Hamiltonian with fixed number of 
electrons.

The VCA accesses the physics of a lattice model in the thermodynamic 
limit by optimizing trial self-energies generated by a reference system.
The reference system consists of an isolated cluster having the same
interaction as the original lattice, but differing in the single-particle 
part of the Hamiltonian. 
An approximation to the self-energy of the lattice model is then obtained 
by finding the saddle point of an appropriate grand-canonical potential
\begin{equation}
\label{omega}
\Omega=\Omega'+ Tr \ln G_{VCA} - Tr \ln G_{cl}
\end{equation}
with respect to the single-particle parameters of the reference 
system.~\cite{pott.03.se} Here, $G_{cl}$ is the Green's function matrix of  
the reference system (cluster), which is calculated numerically as a function 
of frequency and spin, using a zero-temperature Lanczos procedure. The lattice 
Green's function $G_{VCA}$ is obtained by a matrix form of the Dyson equation, 
whereby the self-energy $\Sigma$ of the reference system is used as an 
approximation to the lattice self-energy. Finally, $\Omega'$ is the 
grand-canonical potential of the reference system.

As a reference system for the VCA calculation, we take a cluster of four 
sites located in the $(110)$ plane, and containing two from each inequivalent 
Cr$_1$ and Cr$_2$ sites.~\cite{elsewhere} We have checked that our results 
do not change substantially when using a cluster of $6$ sites.  We stress 
that the remaining hoppings, namely the ones between Cr$_1$-Cr$_1$ 
along $x$ and $y$ are not neglected, but taken into account in the
noninteracting lattice Green's function entering the Dyson equation in
the expression for the VCA Green's function. Quite generally, multi-orbital 
strongly-correlated systems show a competition between different magnetic 
and orbital-ordered phases. In order to check that the ferromagnetic phase 
is the one with the lowest energy, we have compared its grand-canonical 
potential with the one of different magnetic phases containing mixed 
antiferromagnetic and ferromagnetic components in different directions
~\cite{elsewhere}. Notice that the particle density in the physical 
system is in general different from the one of the reference 
system~\cite{ai.ar.06}; the former turns out to be slightly doped with 
$n=1.81$ particles per unit cell. Moreover, for the parameters we have 
used here, which are the appropriate ones for CrO$_2$, we confirm that 
the ferromagnetic state is more favorable energetically with respect to 
all antiferromagnetic states we have considered, as well as to the 
paramagnetic state.

\begin{figure}[h]
\includegraphics[width=0.9\columnwidth]{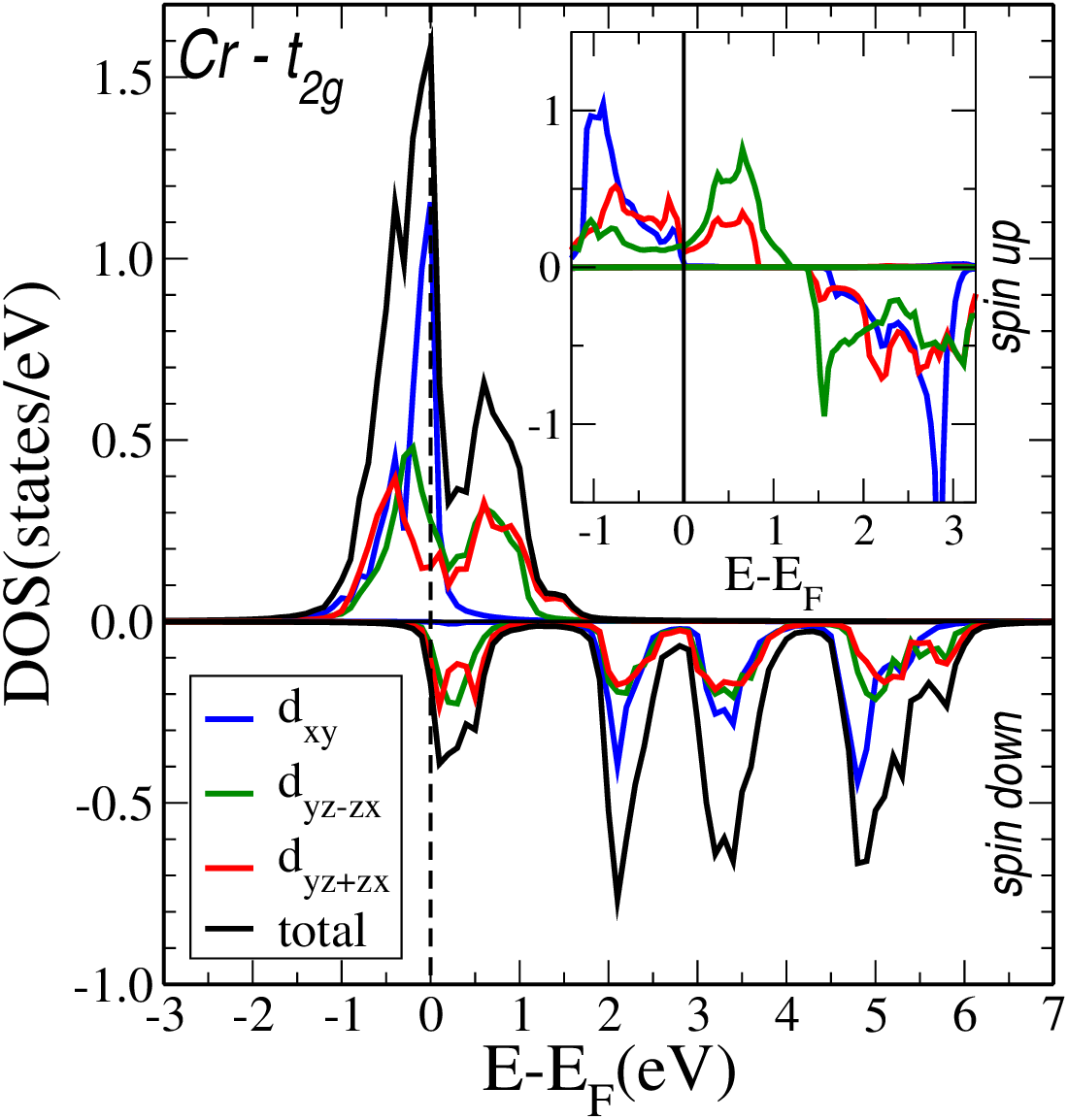}
\caption{(Color online) Cr-t$_{2g}$ orbital-resolved density of states in the 
ferromagnetic state, calculated within the LDA+VCA approach for U=3eV, and J=0.9eV. 
NQP states are visible in the minority spin channel, just above the Fermi 
energy. The inset shows the LSDA+U density of states for the same parameters.}
\label{dos}
\end{figure}

\begin{figure}[h]
\includegraphics[width=0.9\columnwidth]{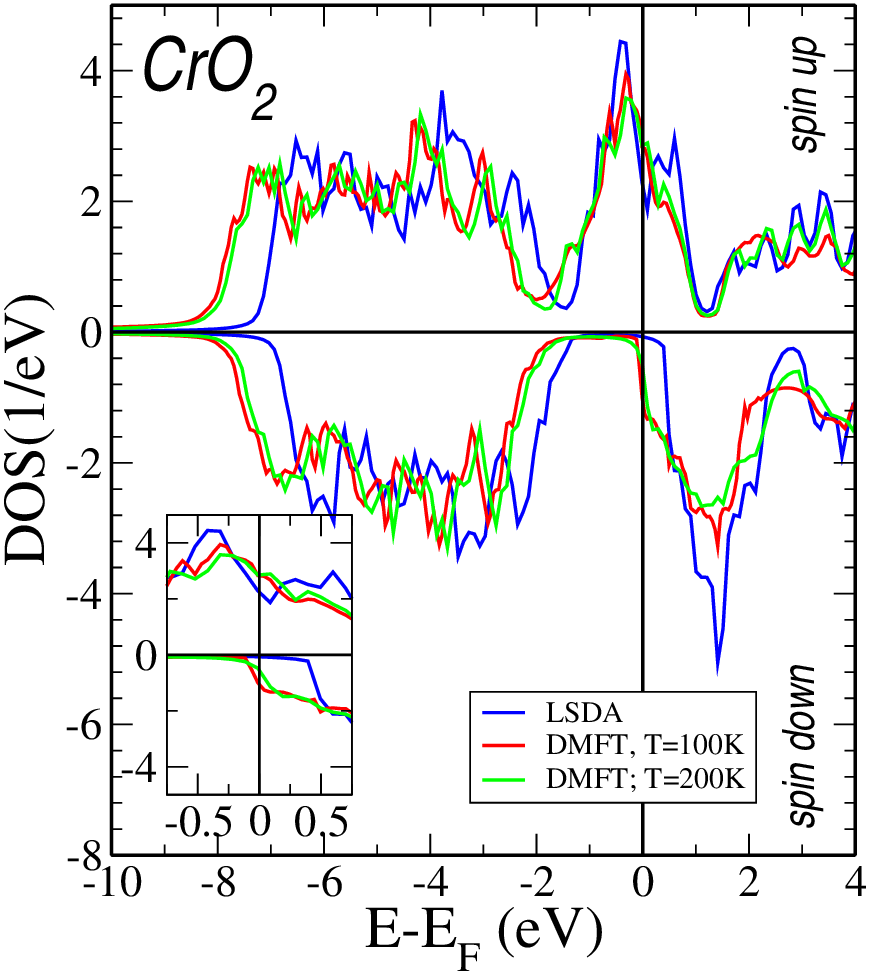}
\caption{(Color online) Density of states obtained within the LSDA and LDSA+DMFT 
calculations for different temperatures. The inset shows the results
for a smaller energy window around the Fermi level.}
\label{dos_cro2}
\end{figure}

In order to analyze the role of the different $d$ orbitals, we present in 
Fig. \ref{dos} the orbital and spin-resolved density of states (DOS). For 
comparison, we show in the inset the results obtained within the mean-field-like 
LDA+U decoupling of the Hubbard interaction for the Cr-t$_{2g}$ 
orbitals.\cite{ko.an.98}
In the LSDA+U, a pseudogap feature opens at the Fermi energy in the majority 
(spin up) DOS, which is attributed to the crystal-field splitting between the 
$d_{yz+zx}$ and $d_{yz-zx}$ orbitals. This pseudogap is already present in the 
(uncorrelated) LSDA results shown in Fig.~\ref{dos_cro2}, although the effects 
of the interaction in LSDA+U enhance this effect. On contrary, in the VCA 
calculation, correlation effects again reduce the pseudogap feature with respect 
to LSDA+U and shift it to higher energies.

A remarkable result of the VCA calculation is the presence of a significant amount 
of density of states at the Fermi energy originating from the putatively localized 
$d_{xy}$ orbital. More specifically, we obtain that this orbital is not completely 
half filled, its occupation being $n^{xy} \approx 0.87$, while for the other two 
orbitals we have $n^{yz + zx} \approx 0.49$, and $n^{yz -  zx}\approx 0.45$. This 
is confirmed by our LSDA+DMFT calculation, discussed below, for which 
$n^{xy}\approx 0.87$, and $n^{yz \pm zx} \approx 0.46$. Consequently, the $d_{xy}$ 
orbital consists of quite itinerant electrons, although with a large effective mass, 
rather than of localized moments. Notice that our findings are in contrast to previous
DMFT \cite{la.cr.01,cr.la.03} calculations, in which  the Fermi energy only touches 
the high-energy tail of the $d_{xy}$ DOS, which can then be considered as localized 
moments. This is probably due to the large value of the interaction parameter $U$ 
used in these calculations. We believe that the smaller value of $U\approx 3eV $ used 
here is more appropriate, as it is obtained from first principles.

In the minority spin channel, NQP states are clearly visible predominantly in the 
$d_{yz \pm zx}$ orbitals just above the Fermi energy. The physical origin of NQP 
states is connected with the ``spin-polaron'' processes \cite{ed.he.73,ir.ka.90}: 
the spin-down low-energy electron excitations, which are forbidden for half-metallic 
ferromagnets in the one-particle picture, turn out to be possible as superpositions 
of spin-up electron excitations and virtual magnons \cite{ed.he.73,ir.ka.90}. 
An uniform superposition forms a state 
with the same total spin quantum number $S=(N+1)/2$ ($N$ is the
particle number in the ground state) as the low-energy spin-up state, but
with one ``spin-flip'', i. e. 
with $z$-component $S_z=S-1$. If the Hamiltonian is spin-rotation invariant
this state with one additional spin-down particle must have the same energy
as the low-energy  state with one additional spin-up particle,
although its weight is reduced by a factor $1/N$.
We stress that spin-rotation invariance is crucial in order to obtain
low-energy NQP states. For this reason, methods neglecting spin flip 
processes in \eeqref{Mu-Hubb1} are not expected to provide a correct 
description of NQP states. Similar NQP states are obtained in the fully 
self-consistent LSDA+DMFT calculation described in the following paragraph. 

The above VCA calculation was carried out by starting from a model
\eref{Mu-Hubb1}, whose parameters are obtained ab-initio from
electronic-structure calculations.
This procedure does not take into account feedback effects of the
correlation-induced charge redistribution into the LSDA potential.
This is  still a difficult task in a VCA calculation.
In addition, correlations and states in the $e_g$ $d$ orbitals are neglected.
In order to explore the consequences of these effects, and  
to access finite temperatures, which is more difficult within a 
Lanczos-VCA approach, we also carried out a multi-orbital 
LSDA+DMFT calculation, which includes the full Cr 
d-manifold, as well as a complete spd-basis set for all the atoms 
in the unit cell. In addition, we include both self-energy and charge 
self-consistency, which means that many-body effects are taken into 
account in the evaluation of the LSDA potentials.
Detailed of this method have been given elsewhere \cite{ch.vi.03}
All previous LSDA+U \cite{ko.an.98, to.ko.05} or DMFT \cite{la.cr.01,cr.la.03} 
studies, independently yield a narrow almost flat band of $d_{xy}$ character 
which produces the ferromagnetic phase of CrO$_2$. We show that in contrast
to these previous results, the fully self-consistent LSDA+DMFT results, in 
agreement with the above non-local VCA approach, yields an itinerant $d_{xy}$ 
orbital, which crosses the Fermi surface. It is interesting to note that despite 
the non-localized nature of the  orbital, we still obtain a ferromagnetic phase.

The DOS obtained from our LSDA+DMFT calculation is presented in Fig.~\ref{dos_cro2} 
for two different values of $T$, and compared with the LSDA results. 
As discussed above, the LSDA Fermi level intersects the majority-spin bands  
near a local minimum and lies in the band gap of the minority spin states. Finite 
temperatures and correlation effects close this minimum around the Fermi 
level, as can be seen from the LDA+DMFT results in Fig.\ref{dos_cro2}. 
No differences can be observed between the two DMFT results at different 
temperatures, except for the smearing of DOS features at higher temperature. 
For both spin channels, the DOS is shifted uniformly to lower energies
in the energy range of -2 and -6eV, where predominantly the $O(p)$ bands 
are situated. This is due to the fact that correlated $Cr(d)$ bands
affect the $O(p)$ states through the $Cr(d)-O(p)$ hybridisation, so that the latter
contribute actively to the mechanism of the ferromagnetic ground state. 
The LSDA+DMFT calculation confirms the existence of minority spin
states just above the Fermi energy, as observed in the VCA calculation.

We now show that many-body effects discussed above, especially the formation of 
NQP states, contribute significantly to the energy dependence of the spin polarization 
\beq
\label{pole}
P(E)=(N_{\uparrow}(E)-N_{\downarrow}(E))/(N_{\uparrow}(E)+N_{\downarrow}(E))
\;,
\eeq
where $N(E)$ is the density of states of majority $\uparrow$ or minority $\downarrow$ 
electrons. 
\begin{figure}[h]
\includegraphics[width=0.9\columnwidth]{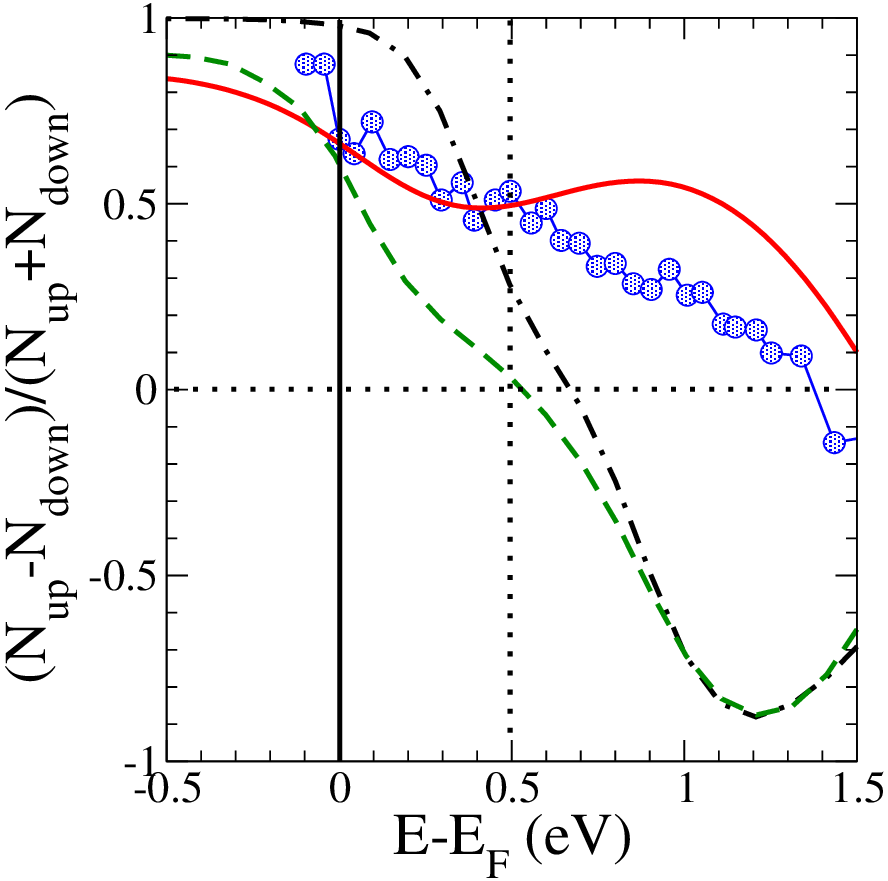}
\caption{(color online)Energy dependence of the spin polarization obtained experimentally
(Ref.~\cite{hu.tj.03} decorated solid line) and by different theoretical calculations
LDA (dot-dashed), DMFT (dashed), VCA (solid) (Eq. \ref{pole}). A broadening 
of 0.4 eV, corresponding to the experimental resolution, has been added to the 
theoretical curves.}
\label{polar_cro2}
\end{figure}
Fig. \ref{polar_cro2} shows a comparison between the measured \cite{hu.tj.03} and 
computed polarization for the different ab-initio many-body calculations discussed 
in the present paper. Due to the tails of NQP states, polarization is less than 
$100\%$  even at the Fermi level. The LDA calculation clearly overestimates the spin 
polarization as it neglects correlation effects. On the other hand, both VCA and DMFT
results show excellent agreement with the experiment at the Fermi level. At an energy 
of about $0.5$ eV polarization is reduced by about $50 \%$, and  up to this energy 
the agreement of the VCA calculation is excellent, while DMFT results overestimate  
depolarisation effects away from the Fermi energy. The disagreement for energies above
$\approx 0.5$ eV  is probably due to terms not included in the VCA Hamiltonian 
\eqref{Mu-Hubb1}, such as $e_g$ orbitals which start becoming important at higher energies.

In conclusion, we addressed the effects of  electronic correlations in CrO$_2$ and 
showed how these considerably change the picture obtained within LSDA and LSDA+U, despite 
the fact that the interaction is not too strong. More specifically, while in LSDA+U exchange 
and crystal-field splitting pin the Cr$d_{xy}$ electrons to become localized moments, the 
competition with charge fluctuation effects, which are taken into account in our DMFT and 
VCA calculations, induce a ferromagnetic state in which also $d_{xy}$ electrons are itinerant 
although with a large effective mass. We believe that this is the correct behavior at this
intermediate value of the interaction parameters, which we derive {\it ab initio},  in contrast 
to results obtained with a larger value of $U$ \cite{ko.an.98,to.ko.05,la.cr.01,cr.la.03}.
We show that correlation-induced NQP states as well as an appropriate description of their 
nonlocal contributions to the self energy as obtained by VCA are crucial for a correct 
description of the energy dependence of the polarization. Additional effects, such as, e. g., 
disorder or phonons, are also expected to contribute to the spin depolarisation.
 
We thank C. Scheiber for help in the spin-rotation invariant part of the code. This work 
is supported by the Austrian science fund (FWF project P18505-N16).   

\bibliographystyle{prsty}
\bibliography{references_database,cro2_dmft}
\end{document}